# Automated crystal orientation mapping by precession electron diffraction assisted four-dimensional scanning transmission electron microscopy (4D-STEM) using a scintillator based CMOS detector

Jiwon Jeong*, Niels Cautaerts, Gerhard Dehm and Christian H. Liebscher

*Department of Structure and Nano-/Micromechanics of Materials, Max-Planck Institut für Eisenforschung GmbH, Düsseldorf, 40237, Germany*

\* Corresponding Author: j.jeong@mpie.de

**Abstract**

The recent development of electron sensitive and pixelated detectors has attracted the use of four-dimensional scanning transmission electron microscopy (4D-STEM). Here, we present a precession electron diffraction assisted 4D-STEM technique for automated orientation mapping using diffraction spot patterns directly captured by an in-column scintillator based complementary metal-oxide-semiconductor (CMOS) detector. We compare the results to a conventional approach, which utilizes a fluorescent screen filmed by an external CCD camera. The high dynamic range and signal-to-noise characteristics of the detector largely improve the image quality of the diffraction patterns, especially the visibility of diffraction spots at high scattering angles. In the orientation maps reconstructed via the template matching process, the CMOS data yields a significant reduction of false indexing and higher reliability compared to the conventional approach. The angular resolution of misorientation measurement could also be improved by masking reflections close to the direct beam. This is because the orientation sensitive, weak and small diffraction spots at high scattering angle are more significant. The results show that fine details such as nanograins, nanotwins and sub-grain boundaries can be resolved with a sub-degree angular resolution which is comparable to orientation mapping using Kikuchi diffraction patterns.

## 1. Introduction

The size and orientation of grains in nanocrystalline materials is directly related to material properties (Meyers et al., 2006). Therefore, the quantitative measurement of local crystallographic orientations is required to understand the structure-property relationship at nanometer scales. For this purpose, several transmission electron microscopy (TEM)-based methods (Schwarzer, 1997) have been developed such as dark-field conical scanning (Li & Williams, 2003; Dingley, 2006; Wu & Zaefferer, 2009), convergent beam electron diffraction (Fundenberger et al., 2003), and nanobeam diffraction (NBD) in scanning transmission electron microscopy (STEM) (Ganesh et al., 2010). Scanning precession electron diffraction (SPED) is also widely used in TEM because the acquisition of diffraction patterns and orientation indexing of each pattern can be carried out automatically which is suitable for automated crystal orientation mapping (Rauch et al., 2008; Viladot et al., 2013; Portillo et al., 2010; Rauch et al., 2010; Cooper et al., 2015).

In SPED, diffraction patterns of individual crystallites are acquired as a four-dimensional (4D) dataset by scanning a focused electron beam in a two-dimensional (2D) array on the sample and (2D) synchronized diffraction patterns are recorded at each probe position, which is also referred to as 4D-STEM (Ophus, 2019). During beam scanning, the incident beam with a diameter of ~1 nm precessed at a constant angle in a conical hollow surface around the optic axis. By recording diffraction patterns with the incident electron beam in precession, the diffraction spot intensities are integrated with an angular range across different diffraction conditions. As a result, quasi-kinematical conditions are achieved by suppressing dynamical scattering effects and a wider range of reflections is excited, which greatly improves the quantitative interpretability of diffraction patterns (Vincent & Midgley, 1994; Own et al., 2006; Oleynikov et al., 2007; Portillo et al., 2010). An orientation map is reconstructed by indexing each diffraction pattern in the 4D-STEM dataset using a template matching algorithm (Rauch et al., 2010; Rauch & Dupuy, 2005). Based on those advantages, SPED enables phase identification and local crystallographic orientation determination of nanostructured or nanocrystalline materials with a high spatial resolution.

However, orientation indexing is often difficult when the acquired diffraction pattern contains all reflections from superimposed grains along the sample thickness (Rauch & Véron, 2019; Kobler & Kübel, 2017). In addition, detailed features such as sub-grain boundaries are difficult to resolve with SPED because of the limited angular resolution (~1°) of the technique (Morawiec et al., 2014; Zaefferer, 2011). In scanning electron microscopy (SEM), transmission Kikuchi diffraction (TKD) has been used as an alternative method for the characterization of nanomaterials because it has required spatial resolution and high angular resolution based on indexing Kikuchi patterns (Sugar et al., 2020; Jeong et al., 2021; Sneddon et al., 2016; Trimby et al., 2014; Ernould et al., 2020; Liu et al., 2019).

The conventional SPED system cannot utilize the orientation sensitive, weak and small diffraction spots at high scattering angle due to the difficulty of capturing those reflections. In conventional SPED configurations, an external charge-coupled device (CCD) camera is used to capture diffraction patterns from a fluorescent screen during nanobeam scanning (Moeck et al., 2011). When the diffracted electrons collide with a fluorescent screen, a phosphor on the fluorescent screen is excited. This produces emitting visible light proportional to the electron intensities on the fluorescent screen. The external CCD camera captures this light as images with an off-axis geometry, which are passed through additional post processing steps, such as distortion and inclination corrections (Eggeman et al., 2015; Moeck et al., 2011; Yao et al., 2016). Although the CCD camera has a high frame rate (~ 180 frames/s) and high sensitivity for fast mapping (Moeck et al., 2011), acquired patterns contain afterimages from the last several positions of the probe because the fluorescent screen maintains to emit light ~100 ms after a diffraction spot has disappeared (Shionoya & Yen, 1998). Due to the various electron to light and light to electron conversion steps in current data acquisition systems, the signal-to-noise ratio for this technique is suboptimal. Hence, this conventional SPED configuration suffers from limitations that introduce scanning artifacts, obscure the detection of weak reflections and strongly restrict the angular resolution of this spot-based technique in the TEM.

Recently developed detectors are promising for the application in automated crystal orientation mapping because they directly capture diffracted electrons in an on-axis geometry. It enables the acquisition of a diffraction pattern with high quality and improved signal-to-noise ratio. In TEM, recent advances have focused on the development of direct electron detectors (Clough et al., 2014). The high electron sensitivity and readout speed of those detectors has propelled their application in numerous 4D-STEM applications (Ophus, 2019). One type of direct electron detector is a hybrid pixel array detector (PAD) which can be used for capturing high intensity reflections with a high dynamic range (Nord et al., 2020; Yang et al., 2015; Mir et al., 2017a). Recently, MacLaren et al. showed an improvement in phase and orientation indexing reliability using a hybrid direct electron detector (MacLaren et al., 2020). Complementary metal-oxide-semiconductor (CMOS) based detectors exhibit a very high electron sensitivity and fast readout speed, which require low dose conditions because the detector can be damaged due to the very thin electron sensitive volume. Therefore, this type of detector has limited applicability for capturing nanobeam diffraction patterns including the high intensity transmitted beam. This drawback can be overcome by using a scintillator coupled CMOS detector. Although electron sensitivity can be slightly degraded, the dynamic range and capacity of high electron dose can be adapted by optimizing the scintillator. However, orientation mapping using a scintillator coupled CMOS detector was not thoroughly investigated and discussed yet.

In this study, we demonstrate orientation mapping with PED assisted 4D-STEM using a scintillator coupled CMOS detector in TEM. In each nano-beam probe position, a high-quality diffraction pattern is acquired with a simultaneous precession of the electron nanobeam. The scintillator was found to be robust during measurement in a TEM operating at 200 kV. The results are compared to data from the conventional orientation mapping system of an external CCD camera filming a fluorescent screen by collecting orientation maps in the same area of a sample using the two techniques. We first compare the data in terms of the quality of the diffraction pattern and orientation map. The following sections focus on the optimization of the template matching process to improve angular resolution using diffraction spot patterns acquired by the CMOS detector.

## 2. Experimental procedures

*TEM sample preparation*

For quantitative measurement of diffraction pattern images, a thin lamella of single crystal Si was prepared in a Scios 2 focused ion beam (FIB, ThermoFisher). The acceleration voltage of the $Ga^+$ ion beam during FIB milling was 30 kV. To minimize the surface damage layers, a final low-energy cleaning step was performed at 0.5 kV.

For orientation mapping, a thin foil of nanocrystalline Cu-Ag alloy with a thickness of 5 μm was used (Oellers et al., 2020). A disc with a diameter of 3 mm was cut out from the foil using a disc puncher and glued to a Cu single hole grid. For perforation, $Ar^+$ ion milling was performed at 2.5 kV using a PIPS II system (Gatan). Low energy milling was followed at 0.5 kV to minimize surface damage and to extend the electron transparent area.

*PED assisted 4D-STEM data acquisition*

PED was performed in a JEM-2200FS TEM (JEOL) operating at 200 kV and equipped with ASTAR (Nanomegas). The microscope was operated in nanobeam diffraction mode with the smallest spot size (Spot 5) and a condenser aperture size of 10 μm. The probe size was measured as ~ 1 nm in diameter with a convergence angle of 2 mrad. A precession frequency of 100 Hz and a precession angle of 0.5° were applied during the nanobeam scanning.

The overall procedure of the orientation mapping using the conventional configuration and the in-column CMOS system are summarized in Fig. 1, respectively. To compare the data quality between the conventional and CMOS camera systems, PED patterns and datasets were acquired in the same region on the same specimen. The first dataset was acquired using the conventional off-axis system filming the fluorescent screen with a Stingray CCD camera (NanoMegas) (labeled as "Conventional" in Fig. 1). This setup is referred to as the conventional configuration in the remainder of the paper. Each acquired pattern has 144 × 144 pixels with an 8-bit depth. The scan size was 150 × 75 pixels with a step size of 2 nm, yielding a scanning area of 300 × 150 nm$^2$. The second dataset was recorded using an on-axis TemCam-XF416 pixelated CMOS detector (TVIPS) (labeled as "CMOS" in Fig. 1). The Universal scan generator (TVIPS) was used for synchronizing the scanning with the acquisition of diffraction patterns. The step size was ~ 2.7 nm for the same scanning area of 300 × 150 nm$^2$. The diffraction patterns had a pixel size of 2k × 2k (2× hardware binning of the full 4k detector area) with 16-bit depth. The exposure time for the acquisition of each diffraction pattern was 50 ms in both experiments. A camera length of 15 cm was selected for all diffraction patterns.

Orientation indexing of both conventional and CMOS datasets was performed by template matching(Rauch et al., 2010) using the ASTAR software package. Note that all images of the CMOS dataset were binned to 512 × 512 pixels (binning factor of 4) and converted to 8-bit depth (Fig. S1) to apply equivalent orientation indexing because the ASTAR software cannot handle the large CMOS dataset and deal with 16-bit data.

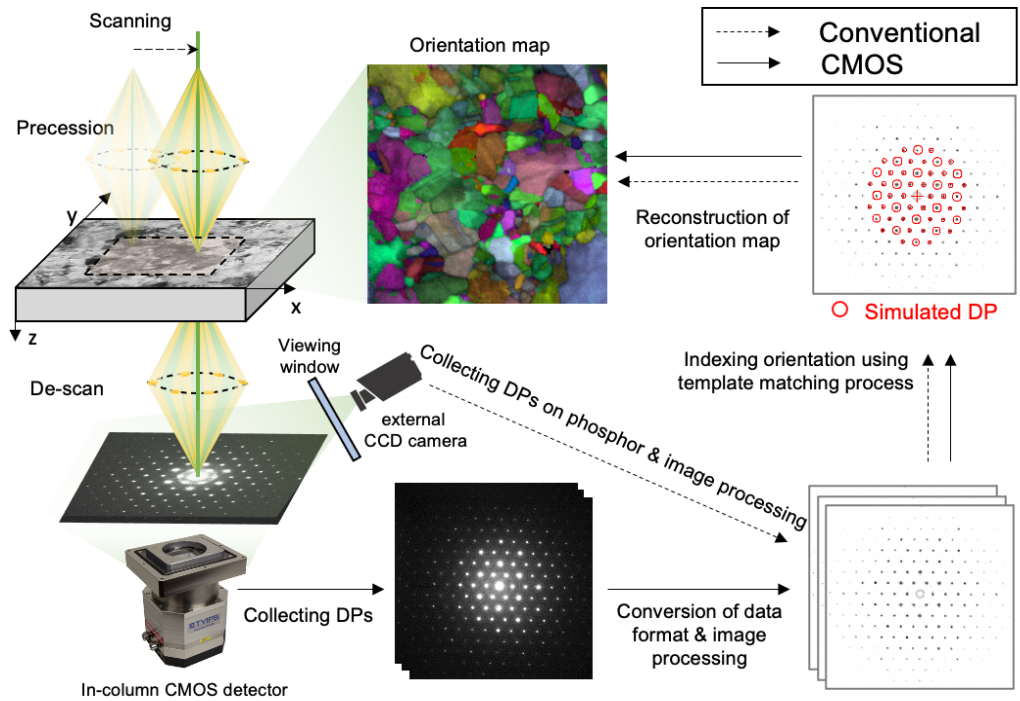

**Figure 1**. Overview of PED orientation mapping analysis using conventional and CMOS camera systems used in the present study.

*Data processing and orientation indexing*

To apply equivalent post-processing, the dataset acquired by the CMOS detector was converted to a block file (.blo format). In this conversion, all images were further binned to 512 × 512 pixels (binning factor of 4) and converted to 8-bit depth using a custom written Python package (Cautaerts, n.d.). Intensity rescaling was also applied to enhance the signal-to-noise ratio of each reflection. The result of this conversion is shown in Fig. S1.

Orientation indexing of both conventional and CMOS datasets was performed by template matching (Rauch et al., 2010) using the ASTAR software package. A library of diffraction templates was generated with a lattice parameter of *a* = 3.615 Å for Cu. The matching index and orientation reliability were calculated for all datapoints during the template matching process as described in the supplementary information (Rauch & Véron, 2014; Rauch & Dupuy, 2005). After orientation indexing, the orientation data was exported (.ang format) and analyzed with the TSL-OIM software (EDAX Inc.). No clean-up process like grain dilation or grain neighboring process was performed for direct comparison.

## 3. Results

*Direct comparison of diffraction patterns acquired by the conventional and CMOS systems*

Figure 2 shows a direct comparison of diffraction patterns obtained by using the two different detector setups. Diffraction patterns were acquired near [110] zone axis of single crystal Si at the same location of the specimen. In the case of the conventional configuration, the original diffraction pattern is stretched along the vertical axis since the fluorescent screen, where the pattern is acquired from with the CCD camera, is inclined with respect to the image plane. After correcting for this inclination, a contraction of the pattern is observed (Fig. 2a). In the CMOS image, more reflections are captured due to the large collection angle of the CMOS detector at the same camera length (Fig. 2b). Distortions in the diffraction pattern acquired using the CMOS detector are mainly stemming from astigmatism and aberrations in the objective and projection lens system. The maximum acceptance angle of each detector was 88 mrad for the conventional configuration and 143 mrad for the CMOS at the camera length of 15 cm, meaning that the acceptance angle of the CMOS detector is 1.625 times larger than that of the conventional system.

Quantitative comparison of intensity profiles extracted along 13 systematic row reflections of two representative nanobeam electron diffraction patterns is shown in Fig. 2c. Since the conventional system has a lower dynamic range, the relative intensities of the transmitted beam and a weak diffracted beam are not remarkable compared to the CMOS detector. Only 8 reflections were captured with the conventional configuration due to the limited acceptance angle and distortion correction (indicated by vertical red lines in Fig. 2c). In contrast, the pattern acquired by the CMOS detector shows a higher number of reflections with a high dynamic range resolved even close to the edge of the detector. In addition, a weak reflection at a high scattering angle (black arrow in Fig. 2c and d) is resolved. This result shows the CMOS detector can capture more weak reflections without the exposure to intensity saturation of the transmitted beam. The radial distribution of the background intensity is observed in both systems which is mainly dominated by inelastically scattered electrons (Fig. 2a and b).

The signal-to-noise ratio for the CMOS detector is also up to ~80 times higher than that for the conventional system (Fig. 2d). The signal-to-noise ratio was calculated as the ratio between the measured intensity and the average intensity of dark noise measured by the conventional (Fig. S2) and the CMOS detector (Fig. S3). The average intensity of dark noise was measured as 18.96 for the external CCD in conventional configuration and 6.07 for the CMOS detector (Fig. S3b), respectively. The signal-to-noise ratio of the conventional configuration is degraded by the conversion efficiency of signal from diffracted electron to light on the fluorescent screen and the detection of reflected light through the viewing window with the external CCD camera (Fig. 1) during data acquisition. Note that the dark noise of the external CCD is amplified due to the image process (gamma correction) which is normally used to compensate for the loss of signal during the suboptimal data acquisition of the conventional system.

While there is a significant difference in image quality between the two methods, orientation determination by template matching is usually successful when the image contains many reflections (e.g. zone axis) in both cases (Fig. 2e and f). In the data from the conventional system, however, artifacts in the form of black horizontal lines are introduced at the pattern edge during image processing (marked by the black arrow in Fig. 2e). This often results in false indexing because the intensity level of these artifacts is comparable to weak reflections.

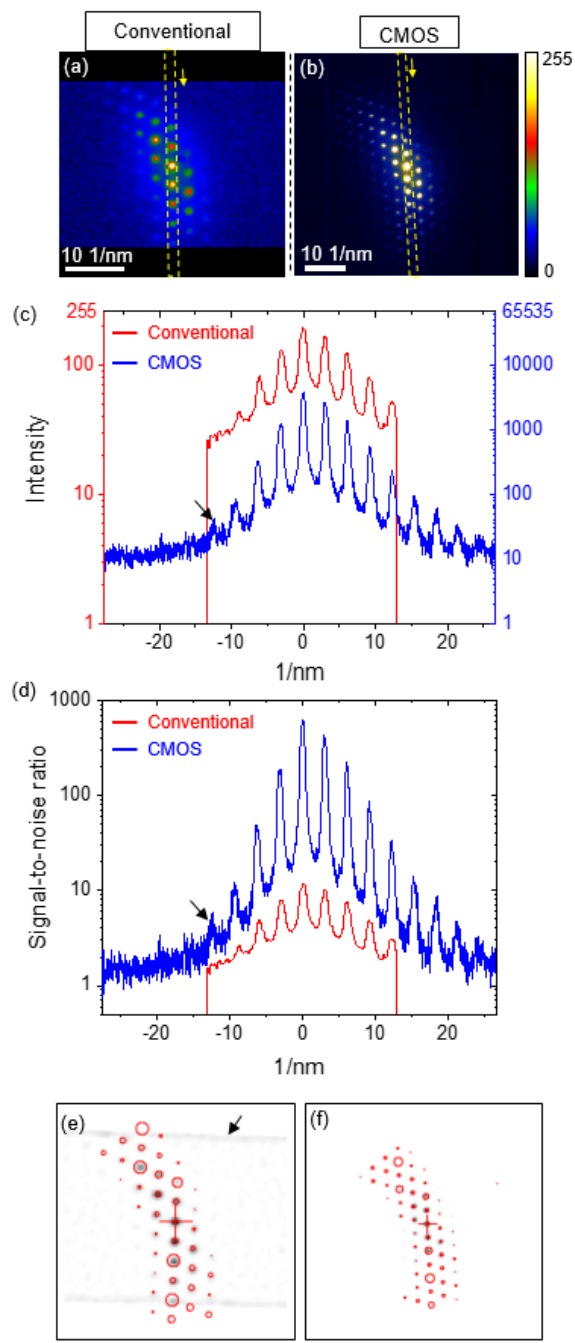

**Figure 2.** Direct comparison of diffraction patterns showing near [110] zone axis of single crystal Si obtained by using conventional and CMOS systems. Raw pattern images with temperature-color scale: (a) conventional and (b) CMOS detector. Note that pixels with an intensity above 255 are shown as white color. (c) Intensity profiles and (d) signal-to-noise ratio measured along the same direction in the PED pattern images (marked as yellow square in Figs. 2a and b). Orientation indexing by template matching process for the pattern images obtained by (e) conventional and (f) CMOS detector. Diffraction spots displayed as red circles represent the simulated pattern with the highest matching index.

*Direct comparison of orientation maps acquired by conventional and CMOS systems*

The orientation maps of the nanocrystalline Cu-Ag alloy acquired by both systems from identical locations of the TEM sample are compared side-by-side in Fig. 3. The bright-field (BF) image (Fig. 3a) and matching index maps (Fig. 3b and c) show nano-sized grains with nanotwins. Σ3 twin boundaries and grain boundaries are represented as yellow (twin), black (high angle grain boundary, misorientation angle >15°), green (low angle grain boundary, 5° ~ 15° misorientation angle) on the orientation maps (Fig. 3d and e). The BF image (Fig. 3a) and raw orientation maps (Fig. 3b-e) are broadly in agreement. However, fine details such as small grains (white arrows in Fig. 3c and e) and nanotwins (yellow arrow in Fig. 3e) can be detected only in the CMOS dataset. Kernel average misorientation (KAM) maps (Fig. 3f and g) show local misorientation within each grain to visualize sub-grain boundary (<5° misorientation). Those boundaries are not well shown in both orientation maps due to the limited orientation resolution. However, more artifacts within the grain are observed in the dataset obtained by the conventional configuration (Fig. 3d and f). Both methods show a reasonable reliability map (Fig. 3h and i) with some dark areas indicating low reliability due to grain overlapping. The orientation resolution is limited to ~1° (the maximum angular difference between adjacent templates is approximately 1° for a library with 1326 templates) because orientations within 1° of all acquired patterns are only matched by one specific simulated pattern.

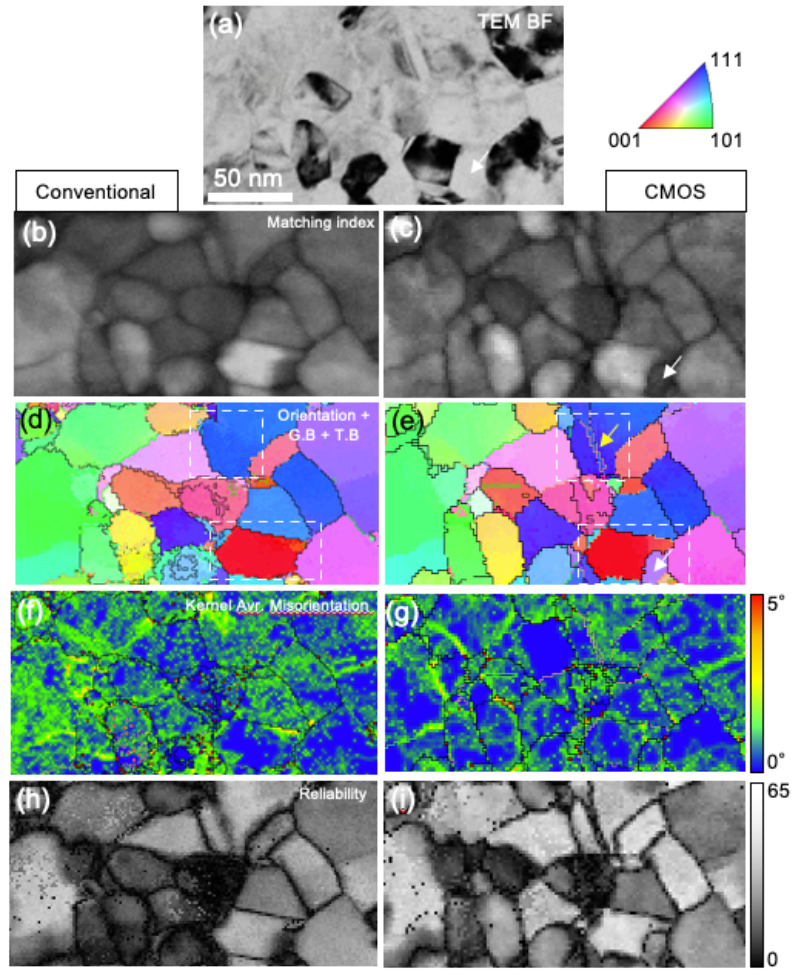

**Figure 3.** Side-by-side comparison of orientation maps of nanocrystalline Cu-Ag sample acquired by conventional configuration and CMOS detector. 1326 simulated patterns were used for the template matching process. (a) BF image, (b, c) matching index maps and (d, e) Orientation maps. Σ3 twin boundary and grain boundaries are represented as yellow (twin), black (high angle grain boundary, misorientation angle >15°), green (low angle grain boundary, 5° ~ 15° misorientation angle). (f, g) Kernel average misorientation maps. Sub-grain boundaries (0° ~ 5° misorientation angle) are displayed as a color scale. The step size was 2 nm for conventional and ~2.7 nm for CMOS.

A detailed comparison of the grain morphology and corresponding pattern indexing of the conventional and CMOS datasets acquired at the same sample regions are shown in Figs. 4 and 5, respectively. Both figures show a BF image of the scanned region, the magnified orientation maps (white dotted square in Figs. 3d and e) and corresponding diffraction patterns. Figure 4 shows the orientation mapping analysis for a grain containing a nanotwin with a thickness of ~ 10 nm (Fig. 4a). While the nanotwin is not resolved in the conventional dataset (Fig. 4b), it is clearly shown with a thickness of 1~2 pixels in the CMOS detector case (Fig. 4d). In the diffraction patterns acquired at the scan position of the nanotwin (Fig. 4c and e), the diffraction patterns of matrix and twin are overlapping with similar intensity. Only a few reflections are visible since the beam direction is away from the zone-axis condition. Due to the advantage of the high dynamic range of the CMOS camera, this nanotwin (Fig. 4d) is resolved by the recognition of a slight difference in the pattern intensity between matrix and twin (Fig. 4e).

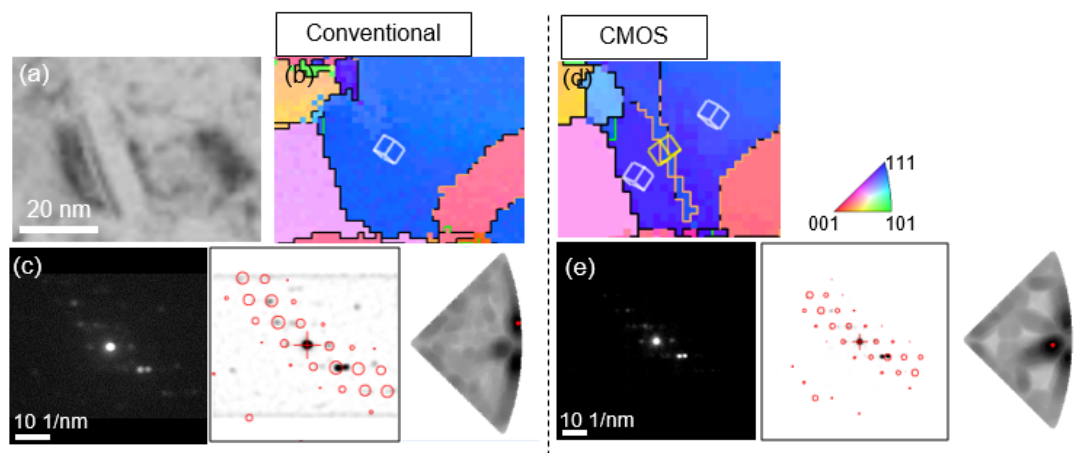

**Figure 4.** Comparison of magnified orientation maps showing the grain containing nanotwin acquired by conventional (Fig. 3d) and CMOS detector (Fig. 3e). (a) BF image. (b, d) Magnified orientation maps. (c, e) Orientation indexing of the nanotwin.

In the conventional case, the grain morphology in the orientation map of Fig. 5b appears elongated along the scanning direction (i.e. left to right) when compared to the BF image shown in Fig. 5a. The diffraction pattern acquired at this grain (marked as a blue circle in Fig. 5b) is matched well with the simulated pattern as shown in Fig. 5c. However, when the beam scans the adjacent grain (marked as a yellow circle in Fig. 5b), the afterimage from the previous beam positions is partially present (white dotted circles in Fig. 5d) in the diffraction pattern of the adjacent grain. As a result, an incorrect orientation is obtained. In the present study, the afterimage was observed along the scanning direction with a maximum of 10 pixels (20 nm with a step size of 2 nm), which results in a higher fraction of mis-indexed data points and a decrease in orientation reliability of the orientation map. By considering an exposure time of 50 ms, the persistence time of the afterimage on the fluorescence screen was estimated to be ~ 500 ms. This problem is notably observed when the beam is scanning from grains close to zone axis orientation to grains oriented away from the zone axis or from the sample to vacuum (Viladot et al., 2013). In the CMOS camera data, the grain morphology in the orientation map (Fig. 5e) more closely resembles the BF image (Fig. 5a) because there is no afterimage present. Hence, diffraction patterns from two nearby grains can be easily distinguished (Fig. 5f and g).

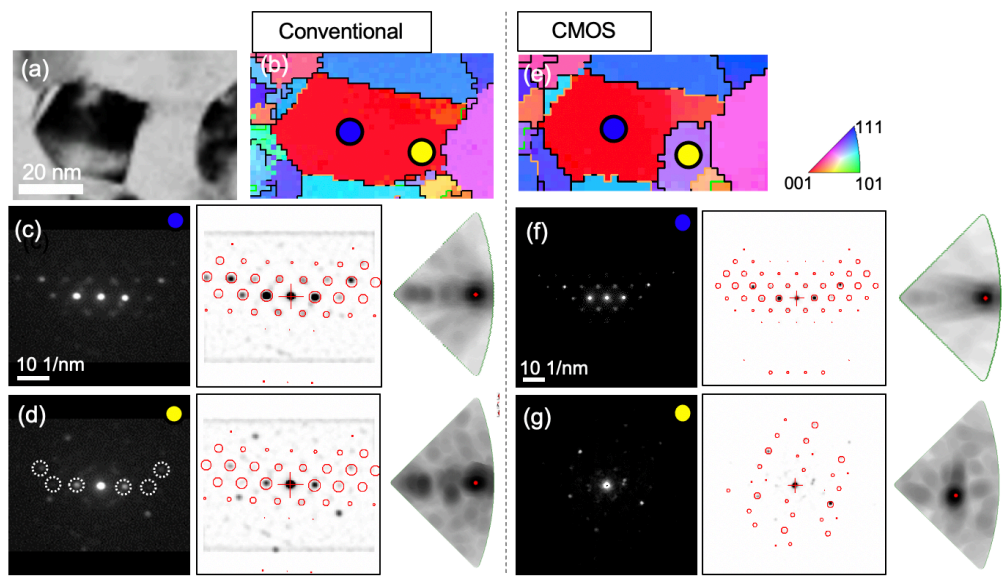

**Figure 5.** Comparison of magnified orientation maps showing nano-sized grains acquired by conventional (Fig. 3d) and CMOS detector (Fig. 3e). (a) BF image. (b, e) Magnified orientation maps. (c, f) Orientation indexing of the zone-axis oriented grain (marked as a blue circle in b and e). (d, g) Orientation indexing of the adjacent grain (marked as a yellow circle in b and e). Reflections from after image are shown as white dot circles.

*Orientation maps acquired by the CMOS system with improved template matching process*

Although weak reflections were captured by the CMOS detector, the default template matching algorithm cannot utilize weak reflections at high scattering angles. One example is shown in Figure 6a. Weak reflections near the detector edge were not matched with the simulated pattern because the reflections near the transmitted beam dominate in the calculation of the matching index. Since they are not orientation sensitive, the template matching is unreliable with ambiguities (Fig. 6b) despite the presence of weak reflections (black arrows in Fig. 6a). In addition, only 1326 templates could be used for orientation indexing because this problem is more pronounced when a larger number of templates were used (Rauch & Véron, 2014). This makes it difficult to resolve the fine details like sub-grain boundaries because the orientation resolution is limited to 1°.

In order to solve this problem, we used a fuzzy mask to exclude reflections near the transmitted beam (dotted circle in Fig. 6c) in the experimental patterns. As a result, only weak and small reflections at high scattering angles (>26 mrad) were considered for the template matching process. Even from a visual inspection a better match of the experimental and simulated patterns is obtained (Fig. 6c). As those reflections are highly orientation sensitive, the reliability of this matching was significantly improved from 8 to 25 with a unique solution (Fig. 6d) while the matching index is decreased from 721 to 305 because the contribution of reflections near the transmitted beam was removed. Note that 11476 simulated patterns were used for orientation indexing with a maximum angular difference between adjacent templates of approximately 0.33°.

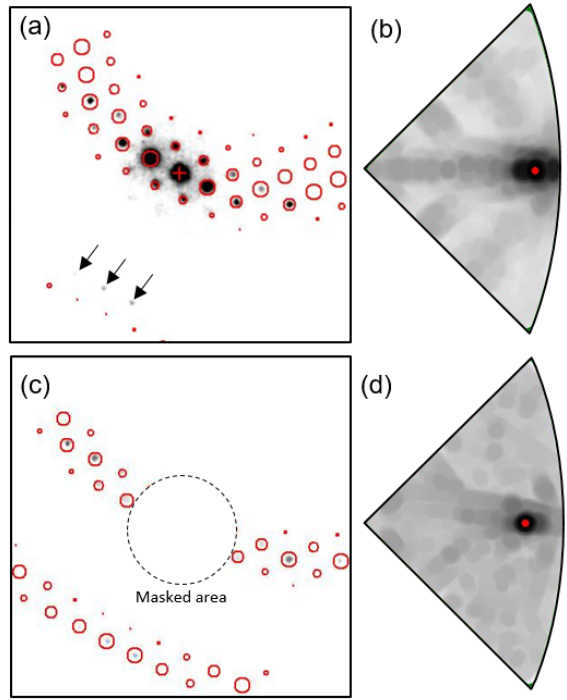

**Figure 6.** Orientation indexing by template matching process for the pattern image (a, b) without masking and (c, d) with masking of reflections near transmitted beam (<26 mrad). The masked area is displayed as a black dotted circle. 11476 simulated patterns were used for orientation indexing. Diffraction spots displayed as red circles represent the simulated pattern with the highest matching index. Orientation is indicated as a red dot in each index map (b and d).

The orientation maps of nanocrystalline Cu-Ag alloy acquired by the CMOS detector using the improved template matching process are shown in Fig. 7. Compared to the orientation map reconstructed with conventional template matching (Fig. 3c and e), the number of artifacts was significantly reduced (Fig. 7b), especially for the detection of grain boundaries with 2° ~ 5° misorientation angle. Sub-grain boundaries within the grain are resolved in the KAM map (Fig. 7c). The reliability map (Fig. 7d) contains pixels with low reliability (low value), especially along the grain boundaries. The main reason is that the template matching process has difficulty finding a unique solution with overlapping patterns at grain boundaries (Rauch & Véron, 2019). In the reliability map, there are some pixels with low reliability inside the grain because the calculation of reliability was incorrect due to the difficulty of finding the two highest local maxima (1$^{st}$ and 2$^{nd}$ possible solution for orientation determination) in matching index map when a large number of simulated patterns were used. Although the calculation of reliability should be optimized further, these results show that the improved template matching process can significantly enhance the angular resolution of spot patterns when orientation sensitive and weak reflections at high scattering angles are detected in the experimental pattern. The precision and accuracy for the misorientation measurement were estimated using orientations of matrix and twin along the Σ3 twin boundary (white arrow in Fig. 7b) which was not resolved with the conventional case (Fig. 3b). Misorientation angles between matrix and twin were measured along the twin boundaries. Euler angles representing the orientation of the matrix and twin are shown in the supplementary material (Table S1). The precision was calculated as the standard deviation of misorientations between matrix and twin. The accuracy was estimated by calculating the difference between ideal misorientation for Σ3 twin (60°) and average measured misorientation between matrix and twin. The angular resolution for misorientation measurement was estimated as 0.20° for precision and 0.27° for accuracy, which is comparable to that of Kikuchi lines (~0.3°) in TEM (Morawiec et al., 2014).

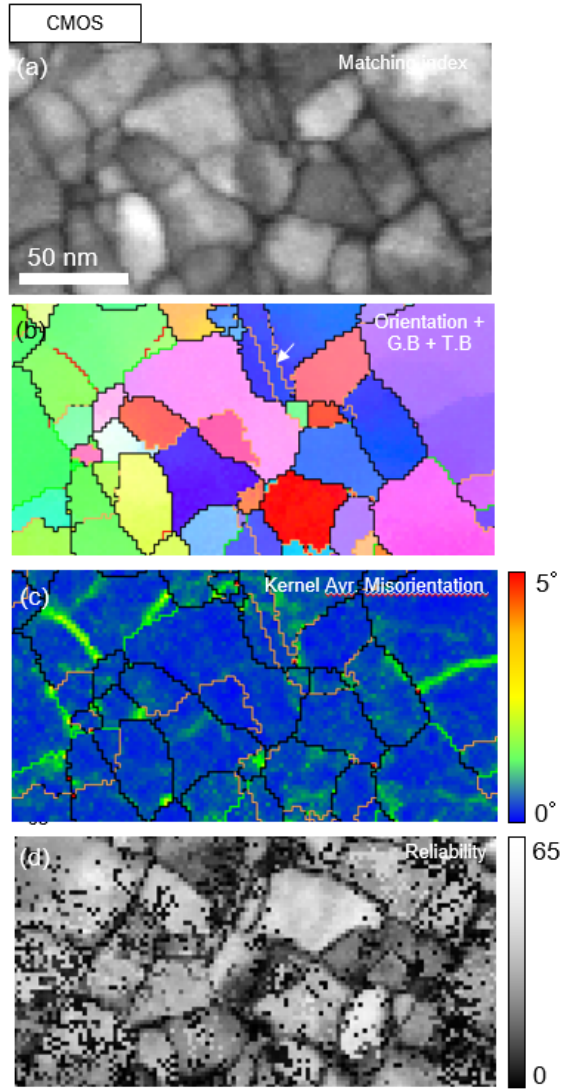

**Figure 7.** Orientation maps of nanocrystalline Cu-Ag sample acquired by conventional and CMOS detector. 11476 simulated patterns (the maximum angular difference between adjacent templates is approximately 0.33°) were used for the improved template matching process (Fig. 6). (a) Matching index map and (b) orientation map with Σ3 twin boundary and grain boundaries represented as yellow (twin), black (high angle grain boundary, misorientation angle >15°), green (low angle grain boundary, 5° ~ 15° misorientation angle), and red lines (2° ~ 5° misorientation angle). (c) Kernel average misorientation map. Sub-grain boundaries (0° ~ 5° misorientation angle) are displayed as a color scale. (d) Reliability map

## 4. Discussion

*Improving the template matching process & angular resolution of the spot pattern*

In the template matching algorithm (Rauch & Dupuy, 2005), simulated patterns contain only position (not radius) and intensity information of each reflection. The correlation between experimental patterns in the 4D-STEM dataset and the simulated patterns in the template library is performed by calculating the matching index which is the sum of products (i.e. the discrete peak intensity values in the simulated pattern times the corresponding measured intensities in the experimental pattern). In simulated patterns, the intensity of reflections at a high scattering angle is relatively high because the atomic scattering factor is not considered (Rauch & Dupuy, 2005) when the pattern was generated, while the intensity of reflections gradually decreases depending on the scattering angle in the experimental pattern (Fig. 2c). This increases the contribution of weak reflections when the matching index is calculated.

However, the current template matching process still could not fully utilize those reflections (Fig. 6a) because orientation indexing is dominated by the high intensity reflections near the transmitted beam. As these reflections are still close to dynamical diffraction condition under precession electron diffraction and their intensity is further contributed by inelastically scattered electrons, this brings them to have the majority in the calculation of matching index regardless of the contribution of weak reflections. In addition, since they have a relatively large spot radius and not orientation sensitive, it can induce mis-indexing of orientation unless the experimental pattern is perfectly corrected in pixels. Therefore, we optimized the template matching process to avoid mis-indexing by applying a fuzzy mask to exclude the reflections near the transmitted beam when the matching index is calculated. As only orientation sensitive and weak reflections are considered, this results in a better agreement between two patterns with an increase of reliability but a decrease of the matching index (Fig. 6).

Orientation mapping with spot patterns in TEM has been shown to have reduced angular resolution compared to Kikuchi-based analysis (Morawiec et al., 2014; Zaefferer, 2011). However, the angular resolution of spot patterns in TEM was not exhausted because the conventional data acquisition system has a limited acceptance angle, limited dynamic range, the low resolution of the diffraction images and low signal-to-noise ratio for detecting weak reflections at high scattering angle. Moreover, although those reflections were captured by the CMOS detector, the current template matching process still has a suboptimal angular resolution. With masked patterns, we could achieve more accurate orientation maps utilizing a large number of templates (high orientation resolution) (Fig. 7).

*The performance of detectors & future remarks*

Since orientation mapping is directly affected by the raw image quality of the diffraction pattern, the performance of detectors and data acquisition process are important to obtain high quality orientation maps. Table 1 shows a comparison between the external CCD camera and recently developed detectors: the scintillator based CMOS detector and the reverse biased hybrid silicon diode PAD. Compared to the conventional case, CMOS and PAD have the advantage of being able to acquire high quality diffraction images due to increased image resolution, and higher dynamic range. Moreover, post-processing of data is much less affected by artifacts because a distortion correction is not required by on-axis data acquisition unless there is astigmatism in the projector system. However, in the conventional case, mis-indexed data points were observed in the orientation map because the pattern image had a low signal-to-noise ratio with limited digital depth (Fig. 2c and d) and suffered from the formation of afterimage (Fig. 5) which could result in incorrect orientation indexing. This might be one reason why a recent study on the comparison between TKD and SPED found a mismatch in the misorientation profiles along the scanning direction beyond stretching or translation that would occur by sample drift (Sugar et al., 2020).

The discrepancy between TKD and SPED measurements might be eliminated or reduced if an on-axis electron detector is used.

It is worth comparing the CMOS detector and hybrid silicon diode PAD. While image resolution is less important for orientation and phase indexing, dynamic range, electron sensitivity and signal-to-noise ratio are crucial parameters. The performance of a detector is quantified by the modulation transfer function (MTF) and detective quantum efficiency (DQE) (Clough et al., 2014). The MTF and DQE are a function of spatial frequency from 0 to 0.5 pixel$^{-1}$. The upper limit is known as the Nyquist frequency, which corresponds to a wavelength of 2 pixels and is the largest possible frequency obtainable by a pixelated image. In most cases, a PAD shows higher performance because diffracted electrons are directly counted and converted as a signal with a larger pixel pitch (55 × 55 μm$^2$) in the detector (direct electron detector). The signal is generated in the active layer and then the electrons exit the active layer before significant lateral scattering occurs (Tate et al., 2016). Scintillator based CMOS systems show a lower performance because diffracted electrons are firstly converted to photons in the scintillator, and then those are detected by the CMOS after photons have passed through the optical fiber bundle for each pixel pitch (15.5 × 15.5 μm$^2$) (indirect electron detector). Therefore, a PAD has an advantage to detect diffracted electrons especially under low dose conditions, because it has a higher electron sensitivity based on the fundamentally different sensor structure.

However, when the acceleration voltage is increased to 200 kV, a fall-off in the MTF and DQE is observed in both detectors. This problem is more pronounced in PAD because higher energy electrons spread further sideways in the Si sensor layer than lower energy ones (Plotkin-Swing et al., 2020; Mir et al., 2017b). This means the performance gap between the two detectors closes with increased acceleration voltage. In the case of DQE, the performance of CMOS is comparable to that of PAD at 200 kV although a CMOS detector has a larger Nyquist limit (Table 1). Furthermore, the background intensity induced by inelastically scattered electrons could be reduced by using an in-column energy filter (Ramachandra et al., 2014). In addition, conventional microscopy such as dark-field imaging and diffraction analysis can be performed using the same detector while the previous study required the additional installation of a direct electron detector for a dedicated SPED experiment (MacLaren et al., 2020; Moeck et al., 2011).

Following the significant improvement of orientation analysis shown in the present study, the on-axis CMOS system has further potential to be used for other applications. In practice, the highly binned diffraction patterns are enough to index the orientation (Rouvimov et al., 2009; Rauch et al., 2010; Eggeman et al., 2015). However, if the material consists of multiple phases with similar lattice parameters, using the CMOS detector at full resolution may be useful to provide an even more reliable orientation map, because slight differences in the positions of reflections could be resolved. Following a similar argumentation, in strain measurements, a higher resolution of the diffraction patterns could provide an improved strain resolution. A remaining challenge lies in the analysis of such large scale datasets, which can exceed 1 TB in size (Spurgeon et al., 2020). Currently, several Python-based packages (Chandra R, 2019; de la Peña et al., 2018; Duncan et al., 2020; Savitzky et al., 2019) are being developed to process such a large dataset (Paterson et al., 2020), which will lead to improved treatment of the 4D-STEM orientation data and also hold great potential for developing advanced orientation indexing approaches.

Table 1. Comparison of the specification of detectors for PED assisted 4D-STEM data acquisition at 200 kV. Image resolution and dynamic range can be adapted to a purpose (e.g. easy handling of a dataset).

| Sensor type | Phosphor CCD (Stingray) | Scintillator based CMOS (XF416) | Reverse biased hybrid silicon diode PAD (Merlin, Medipix3) |
|---|---|---|---|
| Mount | External, off-axis | Retractable, on-axis | Retractable, on-axis |
| Electron sensing mechanism | Phosphor (electron → light) | Scintillator (electron → light) | Direct detection of electrons |

|  | to external CCD | to CMOS | |
| :---: | :---: | :---: | :---: |
| Detector size (mm$^2$) | 35 × 35 | 63.5 × 63.5 | 14 × 14 (Merlin) 17.3 × 14.1 (Medipix3) |
| Pixel size (μm$^2$) | 60 × 60 | 15.5 × 15.5 | 55 × 55 |
| Image resolution | 144 × 144 | 4,096 × 4,096 | 256 × 256 |
| Bit Depth | 256 (8-bit) | 65,536 (16-bit)* | 4,096 (12-bit)** |
| MTF (at 200 kV) | - | 4% at Nyquist, 27% at Nyquist/2 | ~10% at Nyquist ~30% at Nyquist/2 |
| DQE (at 200 kV) | - | ~4% at Nyquist ~28% at Nyquist/2 | ~10% at Nyquist ~20% at Nyquist/2 |
| Maximum frame rate (fps) | 180 (8-bit) | 24 @ 4k × 4k (16-bit) 192 @ 512 × 512 (16-bit) | 1,825 (12-bit) |
| Data size (GB, 200 × 200 pixels) | 0.85 | 1,280 @ 4k × 4k 20 @ 512 × 512 | ~3.75 |

*Ratio between maximum intensity and noise is 20000: 1, **Maximum 24 bit - up to 16.7 million counts per pixel

## 5. Conclusion

We demonstrate a precession electron diffraction assisted 4D-STEM technique for orientation mapping using spot patterns captured by a fast, scintillator coupled CMOS detector. While the orientation map obtained by the conventional system shows mis-indexed data points and scanning noise, the data acquisition with the high-resolution CMOS detector strongly suppresses these artifacts because the image quality of the diffraction patterns was improved by reducing both the background intensity and the formation of an afterimage during data acquisition. Moreover, the angular resolution of misorientation measurement could also be improved by optimizing the template matching process with masked diffraction patterns and a large number of templates. Based on the above advantages, fine details such as nano-sized grains, nanotwins and sub-grain boundaries are resolved in the orientation map.

## Acknowledgments

This work was supported by the Max Planck Society. The authors thank Dr. Viswanadh Gowtham Arigela for providing Cu-Ag thin foils. GD acknowledges financial support by the ERC Advanced Grant GB-Correlate (Grant Agreement: 787446).